\documentclass[runningheads]{svmult}

\usepackage{makeidx}   
\usepackage{graphicx}  
\usepackage{subeqnar}  
\usepackage{multicol}  
\usepackage{physprbb}  
\makeindex             

\begin{document}

\title*{Color Superconductivity in Compact Stars\footnote{To
appear in the proceedings of the ECT Workshop on Neutron
Star Interiors, Trento, Italy, June 2000.  Shorter versions
contributed to the proceedings of Strong and Electroweak Matter 2000,
Marseille, France, June 2000
and to the proceedings of Strangeness 2000, Berkeley, CA, July 2000.
KR was the speaker at all three meetings. We 
acknowledge helpful
discussions with P.~Bedaque, J.~Berges, D.~Blaschke,
I.~Bombaci, G.~Carter,
D.~Chakrabarty,
J.~Madsen, C.~Nayak, M.~Prakash, 
D.~Psaltis, S.~Reddy, M.~Ruderman, S.-J.~Rey, 
T.~Sch\"afer, A.~Sedrakian,
E.~Shuryak, E.~Shuster, D.~Son, M.~Stephanov,
N.~Stergioulas, I.~Wasserman, F.~Weber
and F.~Wilczek.  KR thanks the organizers all three
meetings named above
for providing stimulating environments, thanks
the organizers of SEWM2K for an excuse to visit Provence
for the first time, and thanks the 
organizers of the Trento workshop for 
providing the venue within which
many of the helpful discussions acknowledged above
took place.
This work is supported in 
part  by the DOE under cooperative research agreement \#DF-FC02-94ER40818.
The work of JB was supported in part by an NDSEG Fellowship; that of 
KR was supported in part by a DOE OJI Award and by the
A. P. Sloan Foundation. Preprint MIT-CTP-3028.}}

\toctitle{Color Superconductivity in Compact Stars}
\titlerunning{Color Superconductivity in Compact Stars}

\author{Mark Alford\inst{1}
\and Jeffrey A. Bowers\inst{2}
\and Krishna Rajagopal\inst{2}}

\institute{Dept. of Physics and Astronomy, University of Glasgow, G12 8QQ,
U.K.
\and Center for Theoretical Physics, Massachusetts Institute
of Technology, Cambridge, MA, USA 02139}

\maketitle

\begin{abstract}
After a brief review of the
phenomena expected in cold dense quark matter, color superconductivity
and color-flavor locking, we sketch some
implications of recent developments
in our understanding of cold dense quark matter for 
the physics of compact stars.  We give a more
detailed summary of our recent
work on crystalline color superconductivity and the consequent
realization that (some) pulsar 
glitches may originate in quark matter.
\end{abstract}

\section{Color Superconductivity and Color-Flavor Locking}

Because QCD is asymptotically free, its high temperature
and high baryon density phases are more simply and more
appropriately described in terms of quarks and gluons
as degrees of freedom, rather than hadrons. The chiral 
symmetry breaking condensate which characterizes
the vacuum melts away. At high temperatures, in the 
resulting quark-gluon plasma phase all of the symmetries
of the QCD Lagrangian are unbroken and the excitations have
the quantum numbers of quarks and gluons. At high
densities, on the other hand, quarks form Cooper pairs
and new condensates develop. The formation of 
such superconducting 
phases~\cite{Barrois,BailinLove,ARW1,RappETC,CFL,Reviews}
requires only weak attractive interactions; these phases may
nevertheless break chiral symmetry~\cite{CFL} and
have excitations with the same quantum numbers as those
in a confined phase~\cite{CFL,SW1,ABR2+1,SW2}.
These cold dense quark matter phases may
arise in the cores of neutron stars; understanding this
region of the QCD phase diagram requires an interplay between
QCD and neutron star phenomenology.

The relevant degrees
of freedom in cold dense quark matter
are those which involve quarks with momenta
near the Fermi surface.  At high density, where the
quark number chemical potential $\mu$ (and hence the quark  
Fermi momentum) is large, the QCD gauge coupling $g(\mu)$ is small.
However, because of the infinite degeneracy among  
pairs of quarks with equal and opposite momenta
at the Fermi surface, 
even an arbitrarily weak attraction between quarks renders
the Fermi surface unstable to the formation of a condensate
of quark Cooper pairs.  
Creating a pair costs no free energy 
at the Fermi surface and the attractive interaction
results in a free energy benefit.
Pairs of quarks cannot be color
singlets, and in QCD with two flavors of massless quarks, they 
form in the (attractive) color ${\bf \bar 3}$ 
channel in which the quarks 
in a Cooper pair are 
color-antisymmetric~\cite{Barrois,BailinLove,ARW1,RappETC}.
The resulting condensate creates a gap $\Delta$ at
the Fermi surfaces of quarks with two out of three colors,
but quarks of the third color remain gapless. Five gluons
get a Meissner mass by the Anderson-Higgs 
mechanism~\cite{TwoFlavorMeissner};
a $SU(2)_{\rm color}$ subgroup remains unbroken.  The Cooper
pairs are flavor singlets and no flavor 
symmetries are broken. There is also an unbroken
global symmetry which plays the role of $U(1)_B$.  
Thus, no global symmetries are broken in this 2SC phase.
There must therefore be a phase transition between the hadronic
and 2SC phases at which chiral symmetry is restored. This
phase transition is first 
order~\cite{ARW1,bergesraj,PisarskiRischke1OPT,CarterDiakonov}
since it involves a competition between chiral condensation
and diquark condensation~\cite{bergesraj,CarterDiakonov}.

In QCD with three flavors of massless quarks, the Cooper
pairs {\it cannot} be flavor singlets, and both color and flavor
symmetries are necessarily broken. The symmetries of
the phase which results have been analyzed 
in~\cite{CFL,SW1}.  The attractive channel favored
by one-gluon exchange exhibits ``color-flavor locking.''
A condensate 
of the form 
\begin{equation}
\label{CFLform}
\langle \psi_L^{\alpha a}\psi_L^{\beta b}\rangle 
\propto \Delta \epsilon^{\alpha\beta A}\epsilon^{abA} 
\end{equation}
involving left-handed quarks alone, 
with $\alpha$, $\beta$ color indices and $a$, $b$ flavor indices,
locks $SU(3)_L$ flavor rotations to $SU(3)_{\rm color}$:
the condensate is not symmetric under either alone, but is
symmetric under the 
simultaneous $SU(3)_{L+{\rm color}}$ rotations.\footnote[1]{It turns
out~\cite{CFL} that condensation in the color ${\bf \bar 3}$ channel
induces a condensate in the color ${\bf 6}$ channel 
because this breaks no further symmetries~\cite{ABR2+1}.
The resulting condensates can be written in terms 
of $\kappa_1$ and $\kappa_2$ where
$\langle \psi^{\alpha a}_L \psi^{\beta b}_L \rangle \sim \kappa_1 
\delta^{\alpha a}
\delta^{\beta b} + \kappa_2 \delta^{\alpha b} \delta^{\beta a}$. Here,
the Kronecker $\delta$'s
lock color and flavor rotations. The pure color ${\bf \bar 3}$ 
condensate (\ref{CFLform}) has $\kappa_2=-\kappa_1$.}
A condensate involving right-handed quarks alone 
locks $SU(3)_R$ flavor rotations to $SU(3)_{\rm color}$.
Because color is vectorial, the combined effect
of the $LL$ and $RR$ condensates is to lock $SU(3)_L$
to $SU(3)_R$, breaking chiral symmetry.\footnote[2]{Once 
chiral symmetry is broken by color-flavor locking, 
there is no symmetry argument precluding the existence
of an ordinary chiral condensate. Indeed,
instanton effects do induce a nonzero $\langle \bar q q \rangle$~\cite{CFL},
but this is a small effect~\cite{RappETC2}.}
Thus, in quark matter with three massless quarks,
the $SU(3)_{\rm color}\times SU(3)_L \times SU(3)_R \times U(1)_B$
symmetry is broken down to the global diagonal $SU(3)_{{\rm color}+L+R}$
group. A gauged $U(1)$ subgroup 
of the original symmetry group --- a linear combination of one
color generator and electromagnetism, which lives within
$SU(3)_L \times SU(3)_R$ --- also remains unbroken.  
All nine quarks have a gap. All eight gluons get a 
mass~\cite{CFL,RischkeMeissner}.
There are nine massless Nambu-Goldstone bosons.
All the quarks, all the massive vector bosons, and
all the Nambu-Goldstone bosons have integer charges
under the unbroken gauged $U(1)$ symmetry, which therefore plays
the role of electromagnetism.
The CFL phase therefore has the same symmetries
(and many similar non-universal features)
as baryonic matter with a condensate of
Cooper pairs of baryons~\cite{SW1}.  
This raises the possibility that quark matter and baryonic
matter may be continuously connected~\cite{SW1}.

Nature chooses two light quarks and one middle-weight
strange quark, rather than three degenerate quarks.
A nonzero $m_s$ weakens condensates which
involve pairing between light and strange quarks.
The CFL phase requires
$\langle us \rangle$ and $\langle ds \rangle$
condensates; such condensates which pair quarks with differing Fermi momenta 
can only exist if the resulting gaps are 
larger than of order $m_s^2/2\mu$, the
difference between $u$ and $s$ Fermi momenta in
the absence of pairing. 
This means that upon increasing $m_s$ at fixed $\mu$, one 
must find a first-order
unlocking transition~\cite{ABR2+1,SW2}: for larger
$m_s$ only $u$ and $d$
quarks pair and the 2SC phase is obtained.  
For
any $m_s\neq \infty$,
the CFL phase is the ground state at arbitrarily
high density~\cite{ABR2+1}.  
For large values of $m_s$,
there is a 2SC interlude: as a function of increasing
$\mu$, one finds a first order phase transition
at which hadronic matter is replaced by quark matter
in the 2SC phase and a subsequent first order phase
transition at a higher $\mu$ above which CFL quark matter
takes over.
For smaller values of $m_s$, the possibility of
quark-hadron continuity~\cite{SW1} arises.\footnote[3]{Note 
that even if the strange and light quarks
are not degenerate,
the CFL phase may be continuous
with a baryonic phase which is dense enough that the Fermi momenta of
all the nucleons and hyperons are comparable; there
must, however, be phase transition(s) between this 
hypernuclear phase and ordinary nuclear matter~\cite{ABR2+1}.}

Much effort has gone into estimating
the magnitude of the gaps in the 2SC and CFL 
phases, and the consequent critical temperature
above which quark matter ceases to be superconducting.
It would be ideal if this task were within the scope of
lattice gauge theory.
Unfortunately, lattice methods relying
on importance sampling have to this
point been rendered exponentially 
impractical at nonzero baryon density by the 
complex action at nonzero $\mu$. There are more 
sophisticated algorithms which
have allowed the simulation of
theories which are simpler than QCD but which
have as severe a fermion sign problem as that in QCD at nonzero 
chemical potential~\cite{MeronCluster}.
This bodes well for the future.\footnote[4]{Note that 
quark pairing can be studied on the lattice in some models
with four-fermion interactions and in two-color QCD~\cite{HandsMorrison}. 
The $N_c=2$ case has also
been studied analytically in Refs.~\cite{RappETC,analytic2color}; 
pairing in this
theory is simpler to analyze because 
quark Cooper pairs are color singlets. 
The $N_c\rightarrow \infty$
limit of QCD is often one in which hard problems become
tractable. However, the ground state of $N_c=\infty$ QCD
is a chiral density wave, not a color superconductor~\cite{DGR}.
At asymptotically high densities 
color superconductivity persists up
to $N_c$'s of order thousands~\cite{ShusterSon,PRWZ} before being
supplanted by the phase described in Ref.~\cite{DGR}.  At any finite
$N_c$, color superconductivity occurs at 
arbitrarily weak coupling whereas
the chiral density wave does not.
For $N_c=3$, color superconductivity is 
still favored over the chiral density wave (although not by much)
even if the interaction 
is so strong that the color superconductivity gap is 
$\sim \mu/2$~\cite{RappCrystal}.
}
Given the current absence of suitable lattice methods, the
magnitude of the gaps in quark matter at large but accessible
$\mu$ has been estimated using two broad strategies.
The first class of estimates are done within the
context of models whose
parameters are chosen to reproduce zero density 
physics~\cite{ARW1,RappETC,bergesraj,CFL,ABR2+1,SW2,CarterDiakonov,RappETC2,Hsu1,SW0,Vanderheyden,RappCrystal}.
The second strategy for estimating gaps and critical
temperatures is to use
$\mu=\infty$ physics as a guide.
At asymptotically large $\mu$, models with short-range interactions
are bound to fail because the dominant interaction is due
to the long-range magnetic interaction coming from single-gluon
exchange~\cite{PisarskiRischke1OPT,Son}.  The collinear infrared
divergence in small angle scattering via one-gluon exchange
(which is regulated by dynamical screening~\cite{Son})
results in a gap which is parametrically larger at $\mu\rightarrow\infty$
than it would be for any point-like four-fermion interaction~\cite{Son}.
Weak coupling estimates of the 
gap~\cite{Son,PisarskiRischke,Hong,HMSW,SW3,rockefeller,Hsu2,ShovWij,SchaeferPatterns,EHHS,BBS,RajagopalShuster,Manuel}
are valid at asymptotically high densities, with 
chemical potentials $\mu\gg 10^8$~MeV~\cite{RajagopalShuster}. 
Neither class of methods can be trusted quantitatively
for quark number chemical potentials $\mu\sim 400-500$~MeV, as
appropriate for the quark matter which may occur
in the cores of neutron stars. It is nevertheless
satisfying that two very different approaches,
one using zero density phenomenology to normalize models, the
other using weak-coupling methods valid at asymptotically
high density, yield predictions for the
gaps and critical temperatures at accessible
densities
which are in good agreement:  
the gaps at the Fermi surface are of order tens to 100~MeV, with
critical temperatures about half as large.

$T_c\sim 50$~MeV is much larger relative to the
Fermi momentum than in 
low temperature superconductivity in metals.
This reflects the fact that color superconductivity
is induced by an attraction due to the primary,
strong, interaction in the theory, rather
than having to rely on much weaker secondary interactions,
as in phonon mediated superconductivity in metals.
Quark matter is a high-$T_c$ superconductor by any reasonable
definition. 
Its $T_c$ is nevertheless low enough that
it is unlikely the phenomenon can be realized in heavy ion
collisions.

\section{Color Superconductivity in Compact Stars}

Our current understanding of the color superconducting
state of quark matter leads us to believe that it
may occur naturally in compact stars. 
The critical temperature $T_c$ below which quark matter 
is a color superconductor is high enough that
any quark matter which occurs within
neutron stars that are more than a few seconds old
is in a color superconducting state.
In the absence of lattice simulations, present theoretical
methods are not accurate enough to determine whether 
neutron star cores are made of hadronic matter or quark
matter.  They also cannot determine whether any quark
matter which arises will be in the CFL or 2SC phase: 
the difference between the $u$, $d$ and $s$ Fermi momenta
will be a few tens of MeV which is comparable to estimates
of the gap $\Delta$; the CFL phase occurs when $\Delta$ is
large compared to all differences between Fermi momenta.
Just as the higher temperature regions of the QCD
phase diagram are being mapped out in heavy ion collisions,
we need to learn how to use neutron star phenomena to 
determine whether they feature cores made of 2SC quark matter,
CFL quark matter or hadronic matter, thus teaching us
about the high density region of the QCD phase diagram.
It is therefore important to look for astrophysical consequences of
color superconductivity.

\subsection{Equation of State} 

Much of the work on the consequences 
of quark matter within a compact star has focussed on
the effects of quark matter on the equation of state,
and hence on the radius of the star.  As a Fermi surface
phenomenon, color superconductivity has little effect on
the equation of state: the pressure is an integral over
the whole Fermi volume.  Color superconductivity 
modifies the equation of state at the $\sim (\Delta/\mu)^2$
level, typically by a few percent~\cite{ARW1}.  Such small effects
can be neglected in present calculations, and for
this reason we will not attempt to survey
the many ways in which observations of neutron stars
are being used to constrain the equation of state~\cite{Henning}.

We will describe one current idea, however.
As a neutron star in a low mass X-ray binary (LMXB)
is spun up by accretion from its companion, it becomes
more oblate and its central density decreases. If it contains
a quark matter core, the volume fraction occupied by this
core decreases, the star expands, and its moment of inertia
increases.  This raises the possibility~\cite{GlendenningWeberSpinup}
of a period during the spin-up history of an LMXB when
the neutron star is gaining angular momentum via accretion,
but is gaining sufficient moment of inertia that its angular
frequency is hardly increasing.  In their modelling of this effect,
Glendenning and Weber~\cite{GlendenningWeberSpinup} discover 
that LMXB's should spend a significant fraction
of their history with a frequency of around 200~Hz,
while their quark cores are being spun out of existence,
before eventually spinning up to higher frequencies.  
This may explain the observation that 
LMXB frequencies are clustered around 250-350~Hz~\cite{vanderKlis},
which is otherwise puzzling in that it is thought that LMXB's provide
the link between canonical pulsars and millisecond pulsars,
which have frequencies as large as 600~Hz~\cite{ChakrabartyMorgan}.
It will be interesting to see how robust the result of 
Ref.~\cite{GlendenningWeberSpinup} is to changes in model
assumptions and also how 
its predictions fare when compared to 
those of other 
explanations which posit upper bounds on LMXB 
frequencies~\cite{Bildsten2},
rather than a most probable frequency range with no 
associated upper bound~\cite{GlendenningWeberSpinup}. 
We note here that because Glendenning
and Weber's  effect depends only 
on the equation of state and not on other
properties of quark matter, the fact that the quark
matter must in fact be a color superconductor
will not affect the results in any significant way.
If Glendenning and Weber's explanation for the observed clustering
of LMXB frequencies proves robust, it would imply
that pulsars with lower rotational frequencies feature quark matter
cores.  

\subsection{Cooling by Neutrino Emission}

We turn now to neutron star phenomena which {\it are} affected
by Fermi surface physics.  For the first $10^{5-6}$ years of its
life, the cooling of a neutron star is governed by the balance
between heat capacity and the loss of heat by neutrino emission.
How are these quantities affected by the presence of a
quark matter core? This has been addressed recently in 
Refs.~\cite{Blaschke,Page}, following earlier work in Ref.~\cite{Schaab}.
Both the specific heat $C_V$  and the neutrino emission rate 
$L_\nu$ are dominated
by physics within $T$ of the Fermi surface.  If, as 
in the CFL phase,  all quarks have a gap $\Delta\gg T$ then
the contribution of quark quasiparticles to $C_V$ and $L_\nu$ 
is suppressed by $\sim \exp(-\Delta/T)$.  There may be other
contributions to $L_\nu$~\cite{Blaschke}, but these are also
very small.  The specific heat is  
dominated by that of the electrons, although it 
may also receive a small contribution from the CFL phase Goldstone
bosons.  Although further work is required, it is already
clear that both $C_V$ and $L_\nu$ are much smaller than in
the nuclear matter outside the quark matter core. This
means that the total heat capacity and
the total neutrino emission rate (and hence
the cooling rate) of a neutron star with a CFL core will 
be determined completely by the nuclear matter outside 
the core.  The quark matter core is ``inert'':
with its small heat capacity and emission rate it
has little influence on the temperature of the star as a whole.
As the rest of the star emits neutrinos and cools, the core
cools by conduction, because the electrons keep it in good thermal
contact with the rest of the star.   These qualitative expectations
are nicely borne out in the calculations presented
by Page et al.~\cite{Page}.

The analysis of the cooling history of a neutron star with 
a quark matter core in the 2SC phase is more complicated.
The red and green up and down quarks pair with a gap
many orders of magnitude larger than the temperature, which is
of order 10 keV, and
are therefore inert as described above.  
Any strange quarks present will form a
$\langle ss \rangle$ condensate with
angular momentum $J=1$ which locks to color
in such a way that rotational invariance is not
broken~\cite{Schaefer1Flavor}.
The resulting gap has been estimated to be of order
hundreds of keV~\cite{Schaefer1Flavor}, although applying
results of Ref.~\cite{BowersLOFF} suggests a somewhat smaller gap, around
10 keV.  The blue up and down quarks also pair, forming
a $J=1$ condensate which breaks rotational invariance~\cite{ARW1}.
The related gap was estimated to be a few keV~\cite{ARW1}, but this 
estimate was not robust and should be revisited in light of more
recent developments given its importance
in the following.  The critical temperature $T_c$ above
which no condensate forms is of order the  zero-temperature gap
$\Delta$. ($T_c=0.57 \Delta$ for $J=0$ condensates~\cite{PisarskiRischke}.) 
Therefore, 
if there are quarks for which $\Delta\sim T$ or smaller, these quarks
do not pair at temperature $T$. Such quark quasiparticles 
will radiate neutrinos rapidly (via direct URCA
reactions like $d\rightarrow u+e+\bar\nu$, 
$u\rightarrow d+e^+ +\nu$, etc.)
and the quark matter core will cool rapidly and determine the
cooling history of the star as a whole~\cite{Schaab,Page}.
The star
will cool rapidly until its interior temperature is
$T<T_c\sim\Delta$, at which time the quark matter core will become
inert and the further cooling history will be dominated by
neutrino emission from the nuclear matter fraction of the star. 
If future data were to show that neutron
stars first cool rapidly (direct URCA) and then cool more
slowly, such data would allow an estimate of the smallest 
quark matter gap. We are unlikely to be so lucky.
The simple observation of rapid cooling would {\it not} be an unambiguous
discovery of quark matter with small gaps; there are other
circumstances in which the direct URCA processes occur.
However, if as data on neutron star temperatures improves in coming
years the standard cooling scenario proves correct,
indicating the absence of the direct URCA processes, 
this {\it would} rule out the presence
of quark matter with gaps in the 10 keV range or smaller.  
The presence of a quark matter core
in which {\it all} gaps are $\gg T$ can never be revealed by
an analysis of the cooling history.

\subsection{Supernova Neutrinos}

We now turn from neutrino emission from a neutron star which
is many years old to that from the protoneutron star 
during the first seconds of  a supernova.
Carter and Reddy~\cite{CarterReddy}
have pointed out that when this protoneutron
star is at its maximum temperature of order 30-50~MeV,
it may have a quark matter core which is too hot for color
superconductivity.  As such a  protoneutron star core cools
over the next few seconds, this quark
matter will cool through $T_c$, entering the color superconducting
regime of the QCD phase diagram.  For $T\sim T_c$, the
specific heat rises and the cooling slows. Then, as $T$ drops
further and $\Delta$ increases to become greater than $T$,
the specific heat drops rapidly. Furthermore, as the number
density of quark quasiparticles becomes suppressed by $\exp(-\Delta/T)$,
the neutrino transport mean free path rapidly 
becomes very long~\cite{CarterReddy}.
This means that all the neutrinos previously trapped
in the now color superconducting
core are able to escape in a sudden burst.  If 
a terrestrial neutrino detector
sees thousands of neutrinos from a future supernova, Carter
and Reddy's results suggest that there may be a signature of the
transition to color superconductivity present in the time distribution
of these neutrinos.  Neutrinos from the core of the protoneutron
star will lose energy as they scatter on their way out, but because
they will be the last to reach the surface of last scattering, they
will be the final neutrinos received at the earth.  If they are released
from the quark matter core in a sudden burst, they may therefore
result in a bump at late times in the temporal distribution of
the detected neutrinos.  More detailed study remains to be done
in order to understand how Carter and Reddy's signature, dramatic
when the neutrinos escape from the core, is processed as the neutrinos
traverse the rest of the protoneutron star and reach their
surface of last scattering.

\subsection{R-mode Instabilities}
  
Another arena in which color superconductivity comes into play
is the physics of r-mode instabilities.  A neutron star whose
angular rotation frequency $\Omega$ is large enough is unstable
to the growth of r-mode oscillations which radiate 
away angular momentum via gravitational waves, reducing $\Omega$.
What does ``large enough'' mean?  The answer depends on 
the damping mechanisms which act to prevent the growth of
the relevant modes.  Both shear viscosity and bulk viscosity
act to damp the r-modes, preventing them from going unstable.
The bulk viscosity and the quark contribution
to the shear viscosity both become exponentially
small in quark matter with $\Delta>T$ and as a result,
as Madsen~\cite{Madsen} has shown, 
a compact star made {\it entirely} of quark matter with
gaps $\Delta=1$~MeV or greater is 
unstable if its spin frequency is greater than tens to 100~Hz.
Many compact stars spin faster than this, and Madsen therefore
argues that compact stars cannot be strange quark stars
unless some quarks remain ungapped.  Alas, this powerful argument
becomes much less powerful in the context of a neutron star
with a quark matter core.  First, the r-mode oscillations 
have a wave form whose amplitude is largest at large radius,
outside the core. Second, in an ordinary neutron star there
is a new source of damping: friction at the boundary between
the crust and the neutron superfluid ``mantle'' keeps the 
r-modes stable regardless of the properties of a quark matter 
core~\cite{Bildsten,Madsen}.

\subsection{Magnetic Field Evolution}

Next, we turn to the physics of magnetic fields within
color superconducting neutron star cores~\cite{Blaschkeflux,ABRflux}.  
The interior
of a conventional neutron star is a superfluid (because of neutron-neutron
pairing) and is an electromagnetic superconductor
(because of proton-proton pairing).  Ordinary magnetic fields
penetrate it only in the cores of magnetic flux tubes.
A color superconductor behaves differently. At first
glance, it seems that because a diquark Cooper pair has nonzero
electric charge, a diquark condensate
must exhibit the standard Meissner effect, expelling
ordinary magnetic fields or restricting them to flux tubes
within whose cores the condensate vanishes.  This is not
the case~\cite{ABRflux}.
In both the 2SC and CFL phase a linear combination
of the $U(1)$ gauge transformation of ordinary electromagnetism
and one (the eighth) color gauge transformation remains unbroken 
even in the presence of the condensate.  This means that 
the ordinary photon $A_\mu$ and the eighth gluon $G_\mu^8$
are replaced by new linear combinations
\begin{eqnarray}
A_\mu^{\tilde Q} &=& \cos\alpha_0 \,A_\mu + \sin\alpha_0\,G_\mu^8
\nonumber\\
A_\mu^{X} &=& -\sin\alpha_0\,A_\mu + \cos\alpha_0\,G_\mu^8
\end{eqnarray}
where $A_\mu^{\tilde Q}$ is massless and $A_\mu^{X}$ is massive.
That is, $B_{\tilde Q}$ satisfies the ordinary Maxwell
equations while $B_X$ experiences a Meissner effect.
The mixing angle $\alpha_0$ is the analogue of the Weinberg
angle in electroweak theory, in which the 
presence of the Higgs condensate causes the $A_\mu^Y$ and the third
$SU(2)_W$ gauge boson to mix to form the photon, $A_\mu$, and 
the massive $Z$ boson.   
$\sin(\alpha_0)$ is proportional to $e/g$ and turns
out to be about $1/20$ in the 2SC phase and $1/40$ in the CFL
phase~\cite{ABRflux}.  This means that the 
$\tilde Q$-photon which propagates in color superconducting
quark matter is mostly photon with only 
a small gluon admixture. If a color superconducting neutron star core 
is subjected to an ordinary magnetic field, it will either
expel the $X$ component of the flux
or restrict it to flux tubes, but it can
(and does~\cite{ABRflux}) admit the great majority of the flux
in the form of a $B_{\tilde Q}$ magnetic field satisfying
Maxwell's equations.   
The decay in time of this ``free field'' (i.e. not in flux tubes) 
is limited by the $\tilde Q$-conductivity of the quark matter.
A color superconductor is not a $\tilde Q$-superconductor --- 
that is the whole point --- but it turns out 
to be a very good
$\tilde Q$-conductor due to the presence of electrons:
the $B_{\tilde Q}$ magnetic field decays only on a time scale 
which is much longer than the age of the universe~\cite{ABRflux}.
This means that a quark matter core within a neutron
star serves as an ``anchor'' for the magnetic field:
whereas in ordinary nuclear matter the magnetic flux
tubes can be dragged outward by the neutron superfluid
vortices as the star spins down~\cite{Srinivasan}, 
the magnetic flux within the 
color superconducting core simply cannot decay.
Even though this distinction is a qualitative one, it
will be 
difficult to confront it with data since what is
observed is the total dipole moment of the neutron star.
A color superconducting
core anchors those magnetic flux lines which pass through
the core, while in a neutron star with no quark matter core
the entire internal magnetic field can decay over time. 
In both cases, however, the total dipole moment can change
since the magnetic flux lines which do not pass through
the core can move.

\section{Crystalline Color Superconductivity and Glitches in Quark Matter}

The final consequence of color superconductivity 
we wish to discuss is the possibility that (some)
glitches may originate within quark matter regions of a 
compact star~\cite{BowersLOFF}.
In any context in which color superconductivity arises
in nature, it is likely to involve pairing between species of quarks
with differing chemical potentials.   
If the chemical potential difference  
is small enough, BCS pairing
occurs as we have been discussing. 
If the Fermi surfaces are too far apart, no pairing between the species is
possible. The transition between the BCS and unpaired states as the
splitting between Fermi momenta increases has been studied in
electron superconductors~\cite{Clogston}, 
nuclear superfluids~\cite{Sedrakian} and QCD
superconductors~\cite{ABR2+1,SW2,Bedaque}, assuming 
that no other state intervenes.  However,
there is good reason to think that another state can occur.  This is
the ``LOFF'' state, first explored by Larkin and Ovchinnikov~\cite{LO}
and Fulde and Ferrell~\cite{FF} in the context of electron
superconductivity in the presence of magnetic impurities.
They found that near the
unpairing transition, 
it is favorable to form a state in
which the Cooper pairs have nonzero momentum. This is favored because
it gives rise to a region of phase space where each of the two quarks
in a pair can be close to its Fermi surface,
and such pairs can be created at low cost in free energy.
Condensates of this sort spontaneously
break translational and rotational invariance, leading to
gaps which vary periodically in a crystalline pattern.
If in some shell within the quark matter core
of a neutron star (or within a strange quark star)  
the quark number densities are
such that crystalline color superconductivity arises,
rotational vortices may be pinned in this shell, making
it a locus for glitch phenomena.

In Ref.~\cite{BowersLOFF}, 
we have explored the range of parameters
for which crystalline color superconductivity occurs in
the QCD phase diagram, upon making various simplifying assumptions.
We focus primarily on a toy model in which the
quarks interact via a
four-fermion interaction
with the quantum numbers of single gluon exchange.
Also, we only consider pairing between $u$ and $d$ quarks, with
$\mu_d=\bar\mu+\delta\mu$ and $\mu_u=\bar\mu-\delta\mu$, whereas
we expect a LOFF state wherever the difference between the Fermi momenta
of any two quark flavors is near an unpairing transition, including,
for example, near the unlocking phase transition between the 2SC and
CFL phases.  

In the LOFF state, each Cooper pair carries 
momentum $2{\bf q}$ with $|{\bf q}|\approx 1.2 \delta\mu$.
The condensate and gap parameter vary in space with wavelength
$\pi/|{\bf q}|$.  In Ref.~\cite{BowersLOFF}, we simplify
the calculation by assuming that the condensate varies in space
like a plane wave, leaving the determination of the crystal
structure of the QCD LOFF phase to future work. 
We give an ansatz for the LOFF wave function,
and by variation obtain a gap equation which allows
us to solve for the gap parameter $\Delta_A$, the free energy and
the values of the diquark condensates which characterize
the LOFF state at a given $\delta\mu$ and $|{\bf q}|$. 
We then vary $|{\bf q}|$, to find the preferred (lowest
free energy) LOFF state at a given $\delta\mu$, and compare
the free energy of the LOFF state to that of the BCS state with
which it competes. We show results for one choice
of parameters\footnote[5]{Our 
model Hamiltonian has two parameters, the four-fermion 
coupling $G$ and a cutoff $\Lambda$.  We 
often use the value of $\Delta_0$, the BCS gap 
obtained at $\delta\mu=0$, to 
describe the strength of the interaction: small $\Delta_0$
corresponds to small $G$.
When we wish to
study the dependence on the cutoff, we vary $\Lambda$
while at the same time varying the coupling $G$ such
that $\Delta_0$ is kept fixed.  We expect that the relation
between other physical quantities and $\Delta_0$ will
be reasonably insensitive to variation of $\Lambda$.} in Fig.~1(a).
The LOFF state is characterized by a gap parameter $\Delta_A$ and a 
diquark condensate, but not by an energy gap in the dispersion
relation: we obtain the quasiparticle dispersion 
relations~\cite{BowersLOFF} and find that they vary
with the direction of the momentum, yielding gaps that vary from zero
up to
a maximum of $\Delta_A$.  The condensate is dominated by
the regions in momentum space in which a quark pair
with total momentum $2{\bf q}$ has both members of
the pair within $\sim \Delta_A$ of their respective 
Fermi surfaces.
\begin{figure}[t]
\begin{center}
\includegraphics[width=2.2in,angle=-90]{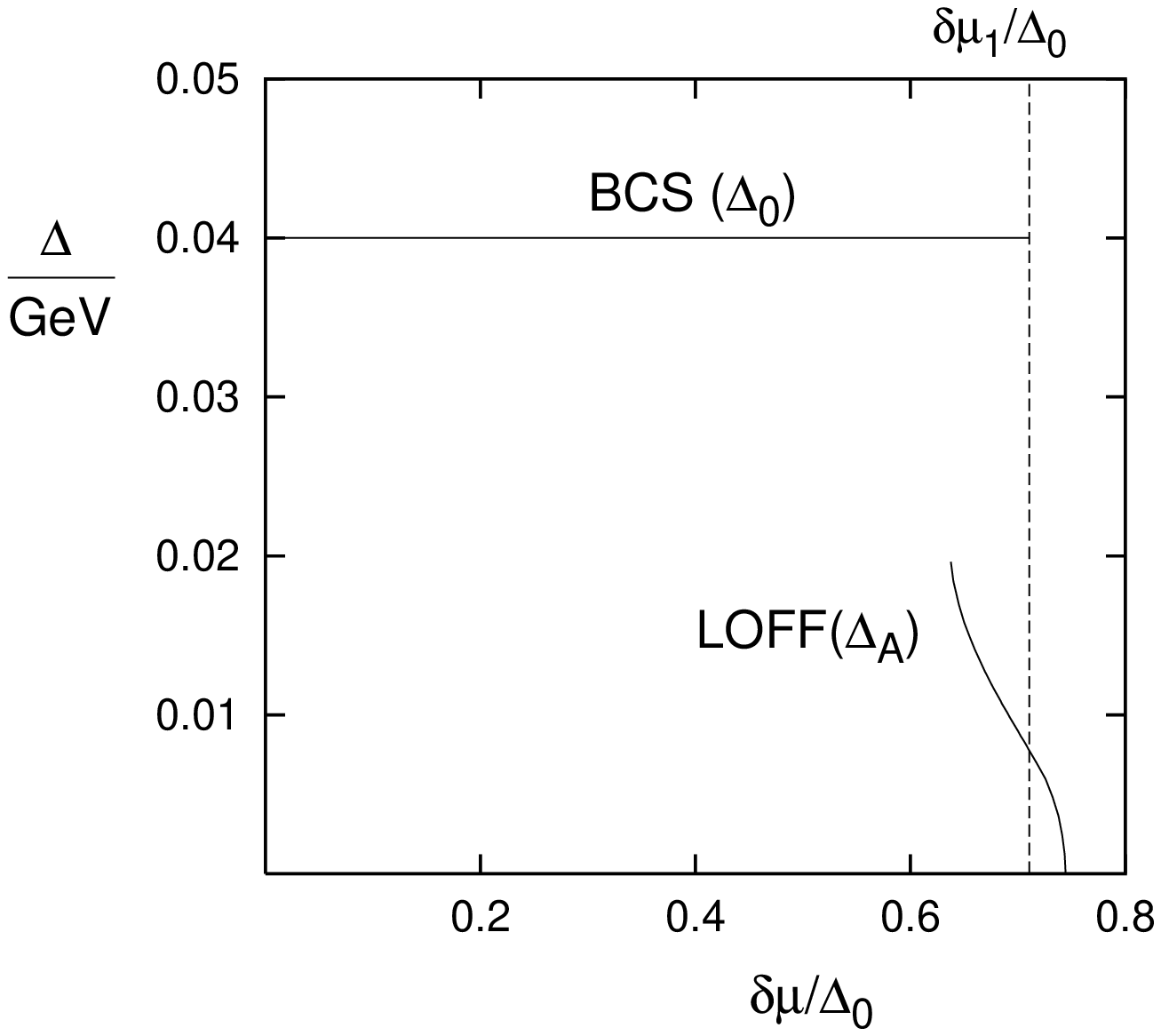}
\includegraphics[width=2.2in,angle=-90]{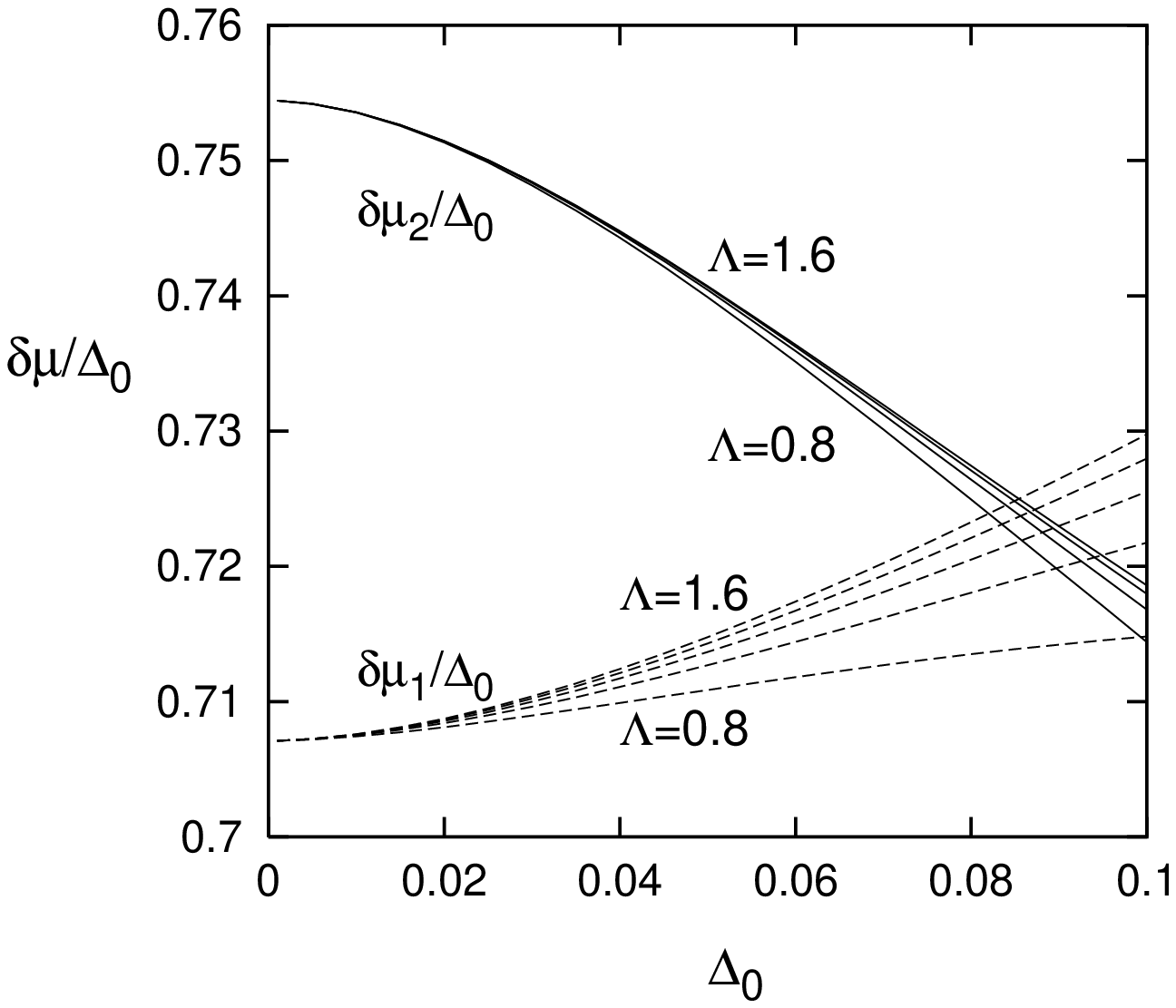}
\end{center}
\caption[]{(a) LOFF and BCS gap parameters as a function of $\delta\mu$,
with coupling chosen so that $\Delta_0=40{\rm ~MeV}$.  At each 
$\delta\mu$ we have varied $|{\bf q}|$ to find the LOFF state
with the lowest free energy.  The vertical dashed line marks 
$\delta\mu=\delta\mu_1$, the value of $\delta\mu$ above
which the LOFF state has lower free energy than BCS.
(b) The interval of $\delta\mu$ within which the LOFF state occurs
as a function of the coupling, 
parametrized by the BCS gap $\Delta_0$ shown in GeV. Here
and in (a), the
average quark chemical potential $\bar\mu$ has been 
set to 0.4~GeV, corresponding to
a baryon density of about 4 to 5 times that in nuclear matter.
A crude estimate~\cite{BowersLOFF} suggests that 
in quark matter at this density, $\delta\mu\sim 15-30$~MeV
depending on the value of the density-dependent effective
strange quark mass.  
Below the solid line, there is a LOFF state. Below the dashed line,
the BCS state is favored. The different lines of each type correspond to
different cutoffs on the four-fermion interaction:
$\Lambda=0.8~{\rm GeV}$ to $1.6~{\rm GeV}$. 
$\delta\mu_1/\Delta_0$ 
and $\delta\mu_2/\Delta_0$ show little
cutoff-dependence, and the cutoff-dependence disappears completely
as $\Delta_0,\delta\mu\rightarrow 0$. $\Lambda=1~{\rm GeV}$ in (a). 
}
\vspace{-0.1in}
\label{fig:fish}
\end{figure}

Because it violates rotational
invariance by involving Cooper pairs whose
momenta are not antiparallel, the quark matter LOFF state  
necessarily features condensates in 
both the $J=0$ and $J=1$ channels. (Cooper pairs in
the symmetric $J=1$ channel 
are antisymmetric in color but symmetric in flavor,
and are impossible in the original LOFF context of pairing 
between electrons, which have neither color nor flavor.)
Both $J=0$ and $J=1$ condensates are present even if there is no
interaction in the $J=1$ channel, as is the
case when we use a four-fermion interaction with the quantum
numbers of Lorentz-invariant single gluon exchange.  Because
there is no interaction in the $J=1$ channel, 
the $J=1$ condensate does not affect the
quasiparticle dispersion relations; that is, the $J=1$
gap parameter vanishes.

The LOFF state is favored
for values of $\delta\mu$ which satisfy 
$\delta\mu_1 < \delta\mu < \delta\mu_2$ as shown in Fig.~1(b),
with $\delta\mu_1/\Delta_0=0.707$ and $\delta\mu_2/\Delta_0=0.754$ in the 
weak coupling limit in 
which $\Delta_0\ll \mu$.  (For $\delta\mu<\delta\mu_1$, we
have the 2SC phase with gap $\Delta_0$.)
At weak coupling, the LOFF gap parameter decreases from $0.23 \Delta_0$
at $\delta\mu=\delta\mu_1$ (where there is a first order BCS-LOFF
phase transition)
to zero at $\delta\mu=\delta\mu_2$ (where there is a second order
LOFF-normal transition).  
Except for very close to $\delta\mu_2$, the critical
temperature above which the LOFF state melts will be much
higher than typical neutron star temperatures.
At stronger coupling the LOFF gap parameter decreases relative
to $\Delta_0$ and 
the window of $\delta\mu/\Delta_0$ within which the LOFF state
is favored shrinks, as seen in Fig.~1(b).  
The single gluon exchange interaction
used in Fig. 1 is neither attractive nor repulsive in
the $J=1$ channel: the width of the LOFF window grows
if the interaction is modified to include an attraction
in this channel~\cite{BowersLOFF}.

Near the second-order critical point $\delta\mu_2$, we can describe the
phase transition with a Ginzburg-Landau effective potential.
The order parameter for the LOFF-to-normal phase transition is
\begin{equation}\label{lofforderparam}
\Phi({\bf r}) = -\frac{1}{2} \langle \epsilon_{ab} \epsilon_{\alpha\beta3} 
\psi^{a\alpha}({\bf r}) C \gamma_5 \psi^{b\beta}({\bf r}) \rangle 
\end{equation}
so that in the normal phase $\Phi({\bf r}) = 0$, while in the LOFF phase
$\Phi({\bf r}) = \Gamma_A e^{i 2 {\bf q} \cdot {\bf r}}$.  (The gap parameter
is related to the order parameter by $\Delta_A=G\Gamma_A$.)
Expressing the order
parameter in terms of its Fourier modes $\tilde\Phi({\bf k})$, we write
the LOFF free energy (relative to the normal state) as
\begin{equation}\label{ginzland}
F(\{\tilde\Phi({\bf k})\}) = \sum_{{\bf k}} \left( C_2(k^2) | 
\tilde\Phi({\bf k}) |^2 
+ C_4(k^2) | \tilde\Phi({\bf k}) |^4 + {\mathcal O}(|\tilde\Phi|^6) \right).
\end{equation}
For $\delta\mu > \delta\mu_2$, 
all of the $C_2(k^2)$ are positive and the normal
state is stable.  Just below the critical point, all of the modes
$\tilde\Phi({\bf k})$ are stable except those on the sphere $|{\bf k}| =
2q_2$, where $q_2$ is the value of $|{\bf q}|$ at $\delta\mu_2$ 
(so that $q_2\simeq 1.2 \delta\mu_2 \simeq 0.9 \Delta_0$ 
at weak coupling).  In general,
many modes on this sphere can become nonzero, giving a
condensate with a complex crystal structure.  We consider the simplest
case of a plane wave condensate where only the one mode
$\tilde\Phi({\bf k} = 2{\bf q}_2) = \Gamma_A$ is nonvanishing.  Dropping all
other modes, we have
\begin{equation}\label{ginzland2}
F(\Gamma_A) = a(\delta\mu - \delta\mu_2) (\Gamma_A)^2 + b (\Gamma_A)^4 
\end{equation}
where $a$ and $b$ are positive constants.  Finding the minimum-energy
solution for $\delta\mu < \delta\mu_2$, we obtain simple power-law relations
for the condensate and the free energy:
\begin{equation}\label{powerlaws}
\Gamma_A(\delta\mu) = K_{\Gamma} (\delta\mu_2 - \delta\mu)^{1/2}, 
\hspace{0.3in} F(\delta\mu) = - K_F ( \delta\mu_2- \delta\mu)^2.
\end{equation}
These expressions agree well with the numerical results obtained
by solving the gap equation~\cite{BowersLOFF}.  
The Ginzburg-Landau
method does not specify the proportionality factors $K_\Gamma$ and
$K_F$, but analytical expressions for these coefficients can be
obtained in the weak coupling limit by explicitly solving the gap
equation~\cite{Takada1,BowersLOFF}, yielding
\begin{equation}\label{keqns}
\begin{array}{rclcl}
G K_\Gamma &=& 2 \sqrt{\delta\mu_2} \sqrt{(q_2/\delta\mu_2)^2  - 1}  
&\simeq& 1.15 \sqrt{\Delta_0} \\[0.5ex]
K_F &=& (4\bar\mu^2/\pi^2)((q_2/\delta\mu_2)^2-1) &\simeq& 0.178 \bar\mu^2.
\end{array}
\end{equation}
Notice that because $(\delta\mu_2-\delta\mu_1)/\delta\mu_2$ is small, the
power-law relations (\ref{powerlaws}) are a good model of the system
throughout the entire LOFF 
interval $\delta\mu_1 < \delta\mu < \delta\mu_2$ where the
LOFF phase is favored over the BCS phase.  The Ginzburg-Landau
expression (\ref{ginzland2}) gives the free energy of the LOFF phase
near $\delta\mu_2$, but it cannot be used to determine the location
$\delta\mu_1$ of the first-order phase transition where the LOFF window
terminates. (Locating the first-order point requires a comparison of
LOFF and BCS free energies.)

The quark matter which may be 
present within a compact star will be in
the crystalline color superconductor (LOFF) state 
if $\delta\mu/\Delta_0$ is in the requisite range.  
For a reasonable value of $\delta\mu$, say 25~MeV,
this
occurs if the gap $\Delta_0$ which characterizes the uniform
color superconductor present at smaller values of $\delta\mu$ is 
about 40~MeV. This is in the middle of the range of present
estimates.  Both $\delta\mu$ and $\Delta_0$ vary as a function
of density and hence as a function of radius in a compact star.
Although it is too early to make quantitative predictions,
the numbers are such that crystalline color superconducting
quark matter may very well occur in a range of radii within a compact 
star. It is therefore worthwhile to consider the consequences.

Many pulsars have been observed to glitch.  Glitches are sudden
jumps in rotation frequency $\Omega$ which may
be as large as $\Delta\Omega/\Omega\sim 10^{-6}$, but may also
be several orders of magnitude smaller. The frequency of observed
glitches is statistically consistent with the hypothesis that 
all radio pulsars experience glitches~\cite{AlparHo}.
Glitches are thought to originate from interactions
between the rigid neutron star crust, typically somewhat 
more than a kilometer thick, and rotational vortices in a
neutron superfluid. 
The inner kilometer of crust
consists of a crystal lattice of nuclei immersed in 
a neutron superfluid~\cite{NegeleVautherin}.
Because the pulsar is spinning, the neutron superfluid 
(both within the inner crust and deeper inside the star) 
is threaded with
a regular array of rotational vortices.  As the pulsar's spin
gradually slows,
these vortices must gradually move outwards since the rotation frequency
of a superfluid is proportional to the density of vortices. 
Deep within the star, the vortices are free to move outwards.
In the crust, however, the vortices are pinned by their interaction
with the nuclear lattice.  
Models~\cite{GlitchModels} differ
in important respects as to how the stress associated
with pinned vortices is released in a glitch: for example,
the vortices may break and rearrange the crust, or a cluster
of vortices may suddenly overcome the pinning force and 
move macroscopically outward, with
the sudden decrease in the angular momentum
of the superfluid within the crust resulting in a sudden increase
in angular momentum of the rigid crust itself and hence a glitch.
All the models agree that the fundamental requirements
are the presence of rotational vortices in a superfluid 
and the presence
of a rigid structure which impedes the motion of vortices and
which encompasses enough of the volume of the pulsar to contribute
significantly to the total moment of inertia.

Although it is 
premature to draw quantitative conclusions,
it is interesting to speculate that some glitches may originate 
deep within a pulsar which features
a quark matter core, in a region of that core 
which is in
a LOFF crystalline color superconductor phase.
A three flavor analysis is required 
to estimate over what range
of densities LOFF phases may arise, as either 
$\langle ud \rangle$, $\langle us \rangle$ or $\langle ds \rangle$
condensates approach their unpairing transitions.  Comparison
to existing models which describe how $p_F^u$, $p_F^d$ and $p_F^s$
vary within a quark matter core in a neutron star~\cite{Glendenning} 
would then
permit an estimate of how much the LOFF region contributes to
the moment of inertia of the pulsar.  Furthermore, a three 
flavor analysis is required
to determine whether the LOFF
phase is a superfluid.   If the only pairing is between $u$
and $d$ quarks, this 2SC phase is not a superfluid~\cite{ARW1,ABR2+1},
whereas if all three
quarks pair in some way, a superfluid {\it is} 
obtained~\cite{CFL,ABR2+1}.
Henceforth, we suppose  that the LOFF phase is a superfluid, 
which means that if it occurs within a pulsar it will be threaded
by an array of rotational vortices.
It is reasonable to expect that these vortices will
be pinned in a LOFF crystal, in which the
diquark condensate varies periodically in space.
Indeed, one of the suggestions for how to look for a LOFF phase in
terrestrial electron superconductors relies on the fact that
the pinning of magnetic flux tubes (which, like the rotational vortices
of interest to us, have normal cores)
is expected to be much stronger
in a LOFF phase than in a uniform BCS superconductor~\cite{Modler}.

A real calculation of the pinning force experienced by a vortex in a
crystalline color superconductor must await the determination of the
crystal structure of the LOFF phase. We can, however, attempt an order
of magnitude estimate along the same lines as that done by Anderson
and Itoh~\cite{AndersonItoh} for neutron vortices in the inner crust
of a neutron star. In that context, this estimate has since been made
quantitative~\cite{Alpar77,AAPS3,GlitchModels}.  
For one specific choice of parameters~\cite{BowersLOFF}, the LOFF phase
is favored over the normal phase by a free energy 
$F_{\rm LOFF}\sim 5 \times (10 {\rm ~MeV})^4$ 
and the spacing between nodes in the LOFF
crystal is $b=\pi/(2|{\bf q}|)\sim 9$ fm.
The thickness of a rotational vortex is
given by the correlation length $\xi\sim 1/\Delta \sim 25$ fm.  
The pinning energy
is the difference between the energy of a section of vortex of length 
$b$ which is centered on a node of the LOFF crystal vs. one which
is centered on a maximum of the LOFF crystal. It 
is of order $E_p \sim F_{\rm LOFF}\, b^3 \sim 4 {\rm \ MeV}$.
The resulting pinning force per unit length of vortex is of order
$f_p \sim E_p/b^2 \sim  (4 {\rm \ MeV})/(80 {\rm \ fm}^2)$.
A complete calculation will be challenging because
$b<\xi$, and is likely to yield an $f_p$
which is somewhat less than that we have obtained by dimensional 
analysis.
Note that our estimate of $f_p$ is
quite uncertain both because it is
only based on dimensional analysis and because the values
of $\Delta$, $b$ and $F_{\rm LOFF}$ are 
uncertain.  (We have a good understanding of 
all the ratios $\Delta/\Delta_0$, $\delta\mu/\Delta_0$, $q/\Delta_0$ 
and consequently $b\Delta_0$ in the LOFF phase.  It is 
of course the value of the BCS gap $\Delta_0$ which is uncertain.) 
It is premature to compare our crude result 
to the results of serious calculations 
of the pinning of crustal neutron vortices as in 
Refs.~\cite{Alpar77,AAPS3,GlitchModels}.  It is nevertheless
remarkable that they prove to be similar: the pinning
energy of neutron vortices in the inner crust 
is $E_p \approx 1-3  {\rm \ MeV}$
and the pinning force per unit length is
$f_p\approx(1-3 {\rm ~MeV})/(200-400 {\rm ~fm}^2)$.

The reader
may be concerned that a glitch deep within the quark
matter core of a neutron star may not be observable:  the
vortices within the crystalline
color superconductor region suddenly unpin and leap
outward;  this loss of angular momentum is compensated
by a gain in angular momentum 
of the layer outside the
LOFF region; how quickly, then, does this increase
in angular momentum manifest itself at the {\em surface} of
the star as a glitch? 
The important point here is that the rotation of any superfluid
region within which the vortices are able to move freely is
coupled to the rotation of the outer 
crust on very short time scales~\cite{AlparLangerSauls}.
This rapid coupling, due to electron scattering off vortices
and the fact that the electron fluid penetrates throughout the 
star, is usually invoked to explain that the core
nucleon superfluid speeds up quickly after a crustal glitch:
the only long relaxation time is that of the vortices within
the inner crust~\cite{AlparLangerSauls}.
Here, we invoke it to explain that the outer crust speeds
up rapidly after a LOFF glitch has accelerated the quark matter
at the base of the nucleon superfluid. 
After a glitch in the LOFF region, the only
long relaxation times are those of the vortices in the LOFF
region and in the inner crust.

A quantitative theory of glitches originating within
quark matter in a LOFF phase must await further 
calculations, in particular a three flavor analysis and
the determination of the crystal structure of the QCD LOFF phase.
However, our rough estimate of the pinning force on 
rotational vortices in a LOFF region suggests that this force may be 
comparable to that on vortices in the inner crust of a conventional
neutron star.
Perhaps, therefore, glitches occurring in a region of crystalline
color superconducting quark matter may yield similar phenomenology
to those occurring in the inner crust.  This is surely
strong motivation for further investigation.

Perhaps the most interesting consequence of these speculations
arises in the context of compact stars made entirely of 
strange quark matter.  The work of Witten~\cite{Witten}
and Farhi and Jaffe~\cite{FarhiJaffe} raised the possibility
that strange quark matter may be stable relative
to nuclear matter even at zero 
pressure.
If this is the
case it raises the question whether observed compact stars---pulsars,
for example---are strange quark stars~\cite{HZS,AFO} rather than
neutron stars.  
A conventional neutron star may feature
a core made of strange quark matter, as we have been discussing 
above.\footnote[6]{Note that a convincing discovery of
a quark matter core within an otherwise hadronic
neutron star would demonstrate
conclusively that strange quark matter is {\it not} stable
at zero pressure, thus ruling out the existence of strange
quark stars.  It is not possible for 
neutron stars with quark matter cores and strange quark
stars to both be stable.}
Strange quark stars, on the other hand, are made (almost)
entirely of quark
matter with either no hadronic matter content at all or
with a thin crust, of order one hundred meters thick, which contains
no neutron superfluid~\cite{AFO,GlendenningWeber}.
The nuclei in this thin crust
are supported above the quark matter by electrostatic forces;
these forces cannot support a neutron fluid.  Because
of the absence of superfluid neutrons, and because of the thinness of
the crust, no successful models of glitches in the crust
of a strange quark star have been proposed.  
Since pulsars are observed to glitch, the apparent lack of a 
glitch mechanism for strange
quark stars  has been the 
strongest argument that pulsars cannot be strange quark 
stars~\cite{Alpar,OldMadsen,Caldwell}.
This conclusion must now be revisited.  

Madsen's conclusion~\cite{Madsen} that a strange
quark star is prone to r-mode instability due to
the absence of damping must
also be revisited, since the relevant oscillations
may be damped within or at the boundary of a 
crystalline color superconductor region.

The quark 
matter in a strange quark star, should
one exist, would be a color superconductor.
Depending on the mass of the star, the 
quark number densities increase by a factor of about two to ten
in going from the surface to the center~\cite{AFO}. This means
that the chemical potential differences among the three
quarks will vary also, and there could be a range of radii
within which the quark matter is in a crystalline
color superconductor phase.  This raises the 
possibility of glitches in strange quark stars.
Because the
variation in density with radius is gradual, if a shell
of LOFF quark matter exists it need not be particularly thin.
And, we have seen, the pinning forces may be comparable
in magnitude to those in the inner crust of a conventional
neutron star.
It has recently been suggested (for reasons unrelated to our considerations)
that certain accreting compact stars
may be strange quark stars~\cite{Bombaci}, although the
evidence is far from unambiguous~\cite{ChakrabartyPsaltis}.
In contrast, 
it has been thought that, because they glitch,  
conventional radio pulsars cannot be strange
quark stars.  Our work questions this assertion
by raising the possibility that glitches
may originate within a layer of quark matter 
which is in a crystalline color superconducting state.


There has been much recent progress in
our understanding 
of how the presence of color superconducting quark matter
in a compact star would affect five different phenomena:
cooling by neutrino emission, the pattern of the arrival times of 
supernova neutrinos, the evolution of neutron
star magnetic fields, r-mode instabilities and glitches.
Nevertheless, much theoretical work remains to be done before
we can make sharp proposals for which 
astrophysical observations can teach us 
whether compact stars contain quark matter, and if so
whether it is in the 2SC or CFL phase.  


%

\end{document}